# Technical Report: CSVM format for scientific tabular data

## CSVM specification (CSVM-1): concept and design, guidelines, parsers.

Gerome Beyries [a,b,c], Frédéric Rodriguez [a,b,*]

[a] CNRS, Laboratoire de Synthèse et Physico-Chimie de Molécules d'Intérêt Biologique, LSPCMIB, UMR-5068, 118 Route de Narbonne, F-31062 Toulouse Cedex 9, France.
[b] Université de Toulouse, UPS, Laboratoire de Synthèse et Physico-Chimie de Molécules d'Intérêt Biologique, LSPCMIB, 118 route de Narbonne, F-31062 Toulouse Cedex 9, France.
[c] Work completed at locations [a,b].

## Abstract

The CSVM (CSV with metadata) is issued from CSV format and used for storing experimental data, models, specifications. CSVM allows the storage of tabular data with a limited but extensible amount of metadata. This increases the exchange and long term use of RAW data because all information needed to use subsequently the data are included in the CSVM file. Basic CSVM files are readable by current tools (i.e. spreadsheets) for handling tables. Using full possibilities of concept, it is possible to deviate from a strict table and annotate also inside the data block. CSVM file are ASCII files and could provide a template for implementing best practices in handling RAW data, in exchange and normalization, in long term resources, or in collaborative processes. In this document we describe the first (CSVM-1) release of CSVM format.

## Keywords

## Status

This document is a proposal for a file format with some guidelines for the design of software toolkits that can handle it. This format if fully documented and publicly available, it can be used as an Open format.

* *Corresponding author.*
CNRS, Laboratoire de Synthèse et Physico-Chimie de Molécules d'Intérêt Biologique, LSPCMIB, UMR-5068, 118 Route de Narbonne, F-31062 Toulouse Cedex 9, France.
Tel.: þ33 (0) 5 61556486; fax: þ33 (0) 5 61556011.
E-mail address: *Frederic.Rodriguez@univ-tlse3.fr* (F. Rodriguez).

# 1. W&H of CSVM File format

The CSVM format is derived from CSV, the main goal behind CSVM is to have a simple data format usable in a generic manner, as closest as possible from CSV. Why and How is explained in this section.

## 1.1. CSV file format

CSV (Comma Separated Values [1]) is a de facto industry standard to exchange tabular data from spreadsheets or databases. A CSV file is constituted from a header line (first line, for column identification) and data lines. Each column is separated by a character not used in data, i.e. a TAB or a comma. A example of CSV tabular data (separators not shown) below:

*Table 1. - CSV table of vehicles.*

| | | | |
|---|---|---|---|
| "ID" | "MODEL" | "TYPE" | "MANUFACTURER" |
| 24 | Xsara | VTS | Citroen |
| #12 | Civic | Type R | Honda |
| 38 | Clio | RS | Renault |
| 12 | Coupé | 16VT | Fiat |
| 45 | 306 | S16 | Peugeot |

Each column is separated by a particular character, i.e. a TAB. To have a more rich content, some conventions must be used in CSV files [2] (using Unicode characters, XML bindings ...).

## 1.2. History of CSVM design

In the end of 90's, engineers and researchers working in the field of environmental sciences had a lot of conceptual problems with their data sets and particularly RAW data. We launched some investigations in the case of collaborative scientific projects (PEVS[3], GIS-ECOBAG P2 : Zone Atelier Adour-Garonne) in the context of Long Term Ecological Ressources (LTERs). After some time (1999-2002) we were able to define what we wanted as data format and metadata. The result was surprising: we wanted the less metadata as possible, not something like XML, not a RDBMS system but flat files. All we defined was in a full contradiction with community's ideas of this time. Now, working outside the RDBMS model is not an exception for database management systems (cf. NoSQL[4] ). Later the work was extended in other scientific fields (chemistry, bioinformatics, protein and enzyme science) and given a proof of concept for format's in terms of genericity.
The main issues were: 1) how we could aggregate raw data in collaborative projects with actors of different scientific fields and 2) how we could ensure a RAW data storage on long term or not (i.e. waiting to be integrated in a RDBMS) without loose information and *in the same time* we collected data.
A classical dialog exists between people which make databases (and need data to do a modeling work) and people which collect data (and expect to use database to store data in the same time). But the scope of our work was not Agile database techniques, because we worked on directly on RAW data.
RAW data is an extended ensemble relative to data itself. Data sets (i.e. in a RDBMS) are designed and populated to answer a scientific question. On the contrary, RAW data could contain a lot of information without interest for the current scientific work, but nobody could say if this information will be used later.
So we wanted to get out database paradigms and work directly on RAW data. We design methods and a file formats able to produce a view of RAW data in response of a scientific stimulus. This view (table(s) or a collection) should be directly usable to produce or populate RDBMS.

1 Comma Separated Values (CSV) [Wikipedia] - http://en.wikipedia.org/wiki/Comma-separated_values
2 http://www.python.org/dev/peps/pep-0305/ or http://www.creativyst.com/Doc/Articles/CSV/CSV01.htm
3 Programme Environnement Vie et Sociétés du CNRS (PEVS)
4 NoSQL [Wikipedia] - http://en.wikipedia.org/wiki/NoSQL

We wanted also to support collective intelligence needed by collaborative projects. These processes involve sharing RAW data of different scientific fields and entities, in the same time. This was the main issue; we expected that if RAW data was correctly annotated and if these annotations were understood by all actors (scientific field vs. scientific field and scientist vs. IT developers) much of the process could be achieved and implemented.

A solution in our context was to add metadata to data tables. The metadata to be written should be as generic as possible, so we excluded data descriptors related to a scientific field. Descriptors should be related to data itself with nearly no modeling of data. The metadata to be written should be also as short as possible, if writers made a big supplementary work to annotate, the aim could not be reached. The ensemble (data + metadata) should be compatible with current tools used by scientists, engineers, or technicians to write data. We placed the metadata block at the bottom of the file after the data block, so the data table could be edited on a spreadsheet, text editor and we retained a CSV (comma separated values) like format. Additional planned features were:

- to embed limited but extensible metadata.
- to be an ASCII format.
- to be the closest as possible from CSV. One user, using the same spreadsheet or text editor, may easily modify CSV data to produce CSVM data or use only the CSV part of CSVM data.
- to be used for experimental data , but also for data description, specifications ... in other ways store in a same format **:** the data, the results, the model.
- to be an alternative for tabular data stored in XML files.
- to be loadable by our applications without implementing a specific parser code for each application.
- to be usable for relational databases import/export (intermediate storage or conversion).

In the period 2002-2004 the concept was in place, we made the design of format itself and we implemented the first toolkit written in Perl Language. In the period 2004-2010 we worked on a Python toolkit and full applications in scientific fields such as medicinal chemistry, structural bioinformatics, were made.
In the recent period (2010- ) an extension of CSVM is on way. Work is carried out to release a new version (CSVM-2) which some improvements such as support of recursive and embedded data.

## 1.3. CSVM and XML

CSVM is not designed to be a substitute of XML or CSV. But if needed, CSVM could offer a full XML interface: a way is to serialize XML file, and code XML tree topology as a particular CSVM column. The binding of CSVM vs. XML is out of the scope of this document.

## 1.4. CSVM by example

The CSV data of *Table1* is rewritten in CSVM and shown below:

*Table 2. - CSVM table of vehicles.*

| | | | | |
|---|---|---|---|---|
| 24 | Xsara | VTS | Citroen | |
| #12 | Civic | Type R | Honda | |
| 38 | Clio | - | Renault | |
| 12 | Coupé | 16VT | Fiat | |
| 45 | 306 | S16 | Peugeot | |
| | | | | |
| #TITLE | Vehicle data | | | |
| #HEADER | ID | MODEL | TYPE | MANUFACTURER |
| #TYPE | NUMERIC | TEXT | TEXT | TEXT |
| #WIDTH | 50 | 50 | 50 | 50 |

The header line is removed and a metadata block (blue, red, green text) is appended at the end of data (orange text). The metadata block is signed by keywords and lines beginning by a # character (blue, red text). The column titles of previous table are now found in a metadata line beginning by keyword #HEADER.

After the #keyword, the order of columns in metadata block is the same as found in data block, with one cell for the keyword (red, blue text).

### Delimiters, empty cells, empty rows, remarks:

Each cell in data block or metadata block is separated by same delimiters as CSV files, i.e. a TAB or another character. Only the colored cells need to be separated by a given delimiter.
An empty cell is defined, as in a CSV file, by two consecutive delimiters. If any char is found (including a SPACE) the cell is not empty. A good practice is to use a specific character to mark empty cells. In the previous example the character '-' is used to mark an empty cell, then it will be more easy to use a text editor rather than a spreadsheet program.

### A true CSVM file

The CSVM file accepts empty rows between the data and metadata (purple) blocks, or inside data block (after Clio model). Rows in data block beginning by # are not read by a CSVM parser, but could be embedded in the CSVM file (red row). Now we show the same CSVM data but with a '!' character as delimiter and as it could be written in a real file.

*Table 3a. - CSVM file of vehicles.*

```
24!Xsara!VTS!Citroen
#12!Civic!Type R!Honda
38!Clio!-!Renault

12!Coupé!16VT!Fiat
45!306!S16!Peugeot

#TITLE!Vehicle data
#HEADER!ID!MODEL!TYPE!MANUFACTURER
#TYPE!NUMERIC!TEXT!TEXT!TEXT
#WIDTH!50!50!50!50
```

### Annotation of data

The previous file shows that a particular row marked by a # symbol (first character) could be masked to the parser. But it is also possible to use this strategy to add information on a row, given the same example with green rows:

*Table 3b. - CSVM file of vehicles.*

```
24!Xsara!VTS!Citroen
#12!Civic!Type R!Honda
38!Clio!-!Renault
# This model is discontinued
# See http://en.wikipedia.org/wiki/Fiat_Coup%C3%A9 for information
12!Coupé!16VT!Fiat
45!306!S16!Peugeot
```

So CSVM admits two levels of annotation: 1) the *metadata block* and 2) *remarks*: insertion of rows marked by # characters in data block. The *metadata* block is compatible with spreadsheets, *remarks* are not. We stop the development of *remarks* at this level, because this could lead to conceive generic descriptors.

# 2. CSVM-1 specifications

Here some specifications describing the flat file format in the first version. This is a framework for the initial design of CSVM but also allowing evolutions of the format. The later versions will add proposals for the insertion of binary objects in cells, some guidelines for the insertion of ASCII data in cells and the insertion of CSVM tables in cells in order to provide tree like structures.

## 2.1. Metadata keywords

The #TITLE, #HEADER and #TYPE keywords, must be always integrated in a CSVM file. These keywords are the support of genericity.

*Table 4. - Metadata keywords.*

| Keyword | Explanation |
|---|---|
| #TITLE | The title of the table/sheet |
| #HEADER | Columns identifiers/titles |
| #TYPE | Data type of columns |
| #WIDTH | An indicator of quantity of text typically contained in column. |

We have also introduced the #META keyword to add something to the metadata block. This keyword is optional. Parsing CSVM file with our API needs only the four previous keywords and will process #META if this keyword is found in the CSVM file.

## 2.2 The #TITLE keyword

This record is used to store general information about the table and uses one field and one row, typically the title of the table. But the string used as a value for this keyword could be used for anything you want.
If this row is used with a secondary separator to store other column metadata than #HEADER and #TYPE, we recommend to use optional #META keyword rather than #TITLE.

## 2.3 The #HEADER keyword

It is obvious than the #HEADER is the title of a column, as we found in some CSV files with this kind of information. Each #HEADER keyword doesn't need to be surrounded by a delimiter such as simple quote or double quote.

## 2.4. The #TYPE keyword

This keyword is used to give the data types for each column. Giving a CSV file for environmental data:

*Table 5. - Working CSV example (some rows are deleted).*

| Diameter | Density | Nature |
|---|---|---|
| 15 | 3 | Perchis |
| 20 | 4 | Perchis |
| … | … | … |
| 55 | 2 | Fut |
| 60 | 1 | Fut |

In this example, the data of columns Diameter et Density (diameter, density) are number, so numerical data , in the other hand, the column Nature (nature) are textual, in fact character strings, and are tagged as text data. The different data types supported are resumed in the following table:

*Table 6. - Values usable with a #TYPE keyword.*

| Sub value | Type |
|---|---|
| NUMERIC | Numerical data (float or integers). |
| TEXT | character strings (unlimited length) |
| DATE | Date, at format DD/MM/YYYY or MM/DD/YYYY |
| BOOLEAN | Two state data. |

### Case of BOOLEAN columns

In the case of BOOLEAN column, the corresponding data in data block must be codes using 0 (false) or 1 (true). But it is possible to code the same values in a TEXT row as 'y' or 'n' or any string.

### Case of NUMERIC columns

The NUMERIC data is used to store numeric values (as text value) but without making a signature on a float or integer values. We expect that after the CSVM parser as loaded all data, the application could check using the #HEADER value the good numeric format.
If not, alternative exists in CSVM to use INTEGER or FLOAT values for #TYPE keywords. But, remember that a CSVM file is loaded by the parser as a string, a matrix of string, a CSVM object embedding a matrix of string. So for a CSVM specification, the value used in #TYPE is not related to a specific processing, all must be done by the application. The value which can be affected to #TYPE keyword is only at programmer's charge. Despite this possibility we recommend to use only very generic types (NUMERIC, INT, INTEGER, FLOAT, REAL, TEXT; STRING etc). See Parsers section of this document for a short discussion about CSVM types.
All of these values must be integrated in a cell after #TYPE keyword, in the same order than data columns. In the case of previous example, we get:

*Table 7. - Working CSVM example (char | used as separator).*

```
15|3|Perchis|
...|...|...|
60|1|Fut|
#TITLE|Foret du boila
#HEADER|Diameter|Density|Nature
#TYPE|NUMERIC|NUMERIC|TEXT
#WIDTH|50|50|100
```

## 2.5. The #WIDTH keyword

This keyword is used in transformations of CSVM files to other formats usable on web pages, to display tabular data (i.e.: Javascript tables, Javascript lists). In other cases, the entire #WIDTH line can be forgotten or, better, filled with values (0, 10, 50 ..) :

*Table 8. - Unused #WIDTH line(filled with zero values).*

```
#WIDTH|0|0|0
```

The unit (the values) used by #WITH is an arbitrary unit, formerly corresponding to a width of column expressed in pixels. In fact, consider that it is an approximation of the maximum quantity of data (string lengths) contained in column for all lines of the CSVM table (see example below):

*Table 9. - Using #WITH to control fields/columns widths.*

```
...|...|...
John-Andrews Howard-Smith|1|yes

#TITLE|Who was here table
#HEADER|name|was_here|Oral communication
#TYPE|TEXT|BOOLEAN|BOOLEANYN
#WIDTH|100|10|10
```

If you use the same CSVM file with without correct proportions (dummy values or zeros) of #WITH, it could be a signal for parser to calculate the accurate values from column contents. But in all cases for using or not #WIDTH, this row must be found in CSVM file.

## 2.6 The #META keyword by example

The following table is used to encode a small collection of molecules with specific values of #TYPE (IMAGE, LINK). The table uses TAB separator (red arrows) and the columns where #TYPE=LINK uses a secondary separator (the | char). The left part of this cell is used to store the location of URL and the right part a name of this URL.

*Figure 1. – A small collection of chemicals (enzyme inhibitors).*

```
inh	glccerase	sketch/mol-22.png	IFM 5-hydroxymethyl-3,4-dihydroxypiperidine 
isofagomine inhibiteur BglA, 1OIF	sketch/mol-22.skc|mol-22.skc
inh	glccerase	sketch/mol-23.png	NOJ 1-deoxynojirimycin inhibiteur BglA, 1OIM	
sketch/mol-23.skc|mol-23.skc
inh	glccerase	sketch/mol-24.png	G2F 2-deoxy-2-fluoro-glucose substrat BglA, 1OIN 
Forme substrat (libère dinitrophénol) présent sous forme d'intermédiaire réactionnel 
(F-Glucose:GLU351)	sketch/mol-24.skc|mol-24.skc
inh	glccerase	sketch/mol-25.png	IFL isofagomine lactam inhibiteur BglA, 1UZ1	
sketch/mol-25.skc|mol-25.skc	lig/ifl.pdb|ifl.pdb
inh	glccerase	sketch/mol-26.png	OXZ tetrahydrooxazine inhibiteur BglA, 1W3J	
sketch/mol-26.skc|mol-26.skc	lig/oxz.pdb|oxz.pdb
inh	glccerase	sketch/mol-27.png	CTS castanospermine inhibiteur BglA, 2CBU	
sketch/mol-27.skc|mol-27.skc	lig/cts.pdb|cts.pdb
inh	glccerase	sketch/mol-28.png	CGB calystegine B2 inhibiteur BglA, 2CBV	
sketch/mol-28.skc|mol-28.skc	lig/cgb.pdb|cgb.pdb
inh	glccerase	sketch/mol-29.png	GIM glucoimidazole inhibiteur BglA, 2CES	
sketch/mol-29.skc|mol-29.skc	lig/gim.pdb|gim.pdb
inh	glccerase	sketch/mol-30.png	PGI pyridine-6,7,8-triol inhibiteur BglA, 2CET	
sketch/mol-30.skc|mol-30.skc	lig/pgi.pdb|pgi.pdb

#TITLE	buildez.org small molecules
#HEADER	TYPE	SECTOR	IMG	NAME	SKETCH	MOL
#TYPE	TEXT	TEXT	IMAGE	TEXT	LINK	LINK
#WIDTH	50	50	50	50	50	50
#META	yes no yes yes yes yes
```

This table will be processed in a workflow and displayed as a dynamic Javascript table, showing images of molecules, and reactive hyperlinks:

*Figure 2. – JavaScript dynamic table issued from the previous CSVM file.*

| TYPE | IMG | NAME | SKETCH | MOL |
|---|---|---|---|---|
| inh | 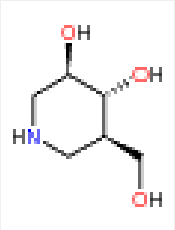 | IFM 5-hydroxymethyl-3,4-dihydroxypiperidine isofagomine inhibiteur BglA, 1OIF | mol-22.skc | ifm.pdb |
| inh | 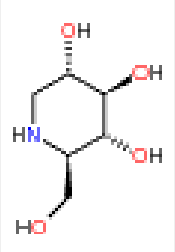 | NOJ 1-deoxynojirimycin inhibiteur BglA, 1OIM | mol-23.skc | noj.pdb |
| inh | 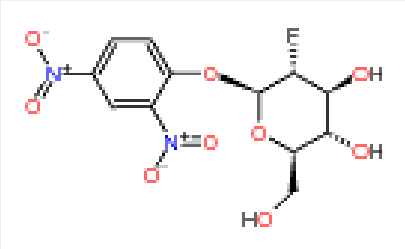 | G2F 2-deoxy-2-fluoro-glucose substrat BglA, 1OIN Forme substrat (libère dinitrophénol) présent sous forme d'intermédiaire réactionnel (F-Glucose:GLU351) | mol-24.skc | g2f.pdb |

Please remark that in this output the column in which #HEADER=SECTOR has disappeared. This particular behavior is coded in #META field of CSVM: 'yes' string is used if column must be displayed and 'no' if not. A SPACE is used as secondary field separator.
This is an example of #META use, this keyword is very often used for storing external reference of data table (i.e. URL, bibliographic references etc).

## 2.7. CSVM Guidelines

Please read carefully information related to CSV format (i.e.: the two links given in first section), CSVM guidelines are simplest that CSV.

### Classical delimiters

Classically, the characters used as CSV delimiters are (in brackets) [, ; : SPACE TAB | !]. Please, take in account that some of these characters can be found in data. So it is better to use delimiter not found in data: as TABs, or rare characters as § (also ALT-245 on PC/Windows). We prefer to avoid some characters found in OS syntax as / \ | or characters such as < >.

### Composite text delimiters

The DOUBLE COTE or SIMPLE COTE characters (" or ') are often used to delimit text data including characters usable also as delimiters (COMMA, SPACES, POINTS ..). The following table gives some examples:

*Table 10. - Using composite text delimiters in CSV files.*

| Data | Delimiter | Number of columns |
|---|---|---|
| "Bonjour, dit'il", c'est pour aujourd'hui ou demain ? | , | 2 |
| Bonjour, dit'il, c'est pour aujourd'hui ou demain ? | , | 3 |
| Bonjour, dit'il, c'est pour aujourd'hui ou demain ? | ' | 4 |
| "Bonjour, dit'il"'c'est pour aujourd'hui ou demain ? | ' | 4 |

Take in account, as shown here, that the corresponding number of columns can be slightly different from one case to other. In order to specify the simplest rules as possible, the CSVM specification don't make use of composite text delimiters ' and ". You must use an adequate column delimiter to keep composite data in each column. A way is to use TABs or '§' character, as shown below:

*Table 11. – CSVM's approach.*

| Data | Delimiter | Number of columns |
|---|---|---|
| Bonjour, dit'il§c'est pour aujourd'hui ou demain ? | § | 2 |

### CSVM and spreadsheets

It is possible to edit a CSVM file using a text editor or a spreadsheet. In this case, after data and metadata integration, you need to export CSVM. This is the same as exporting Text CSV data. If the spreadsheet uses more than one sheet, you cannot export the sheets in the same CSVM file, but each sheet in a separate file. Successfully CSVM export has been done from Open Office and Microsoft software.

### Changing delimiters in CSV/CSVM files

Use a text editor or a word processor and use the Find/Replace tool. A CSVM Python toolkit (Pybuild) provides also functions to do this kind of task.

## 3. Using CSVM with Perl programming language

The following scheme explains in 3 main layers, how we could use CSVM data. The bottom layer is the CSVM file itself, the CSVM API is used to read the file and convert data in a CSVM object including a data matrix (or a data matrix directly). Then the upper layer (at application level) uses the CSVM object.

*Figure 3. - CSVM model.*

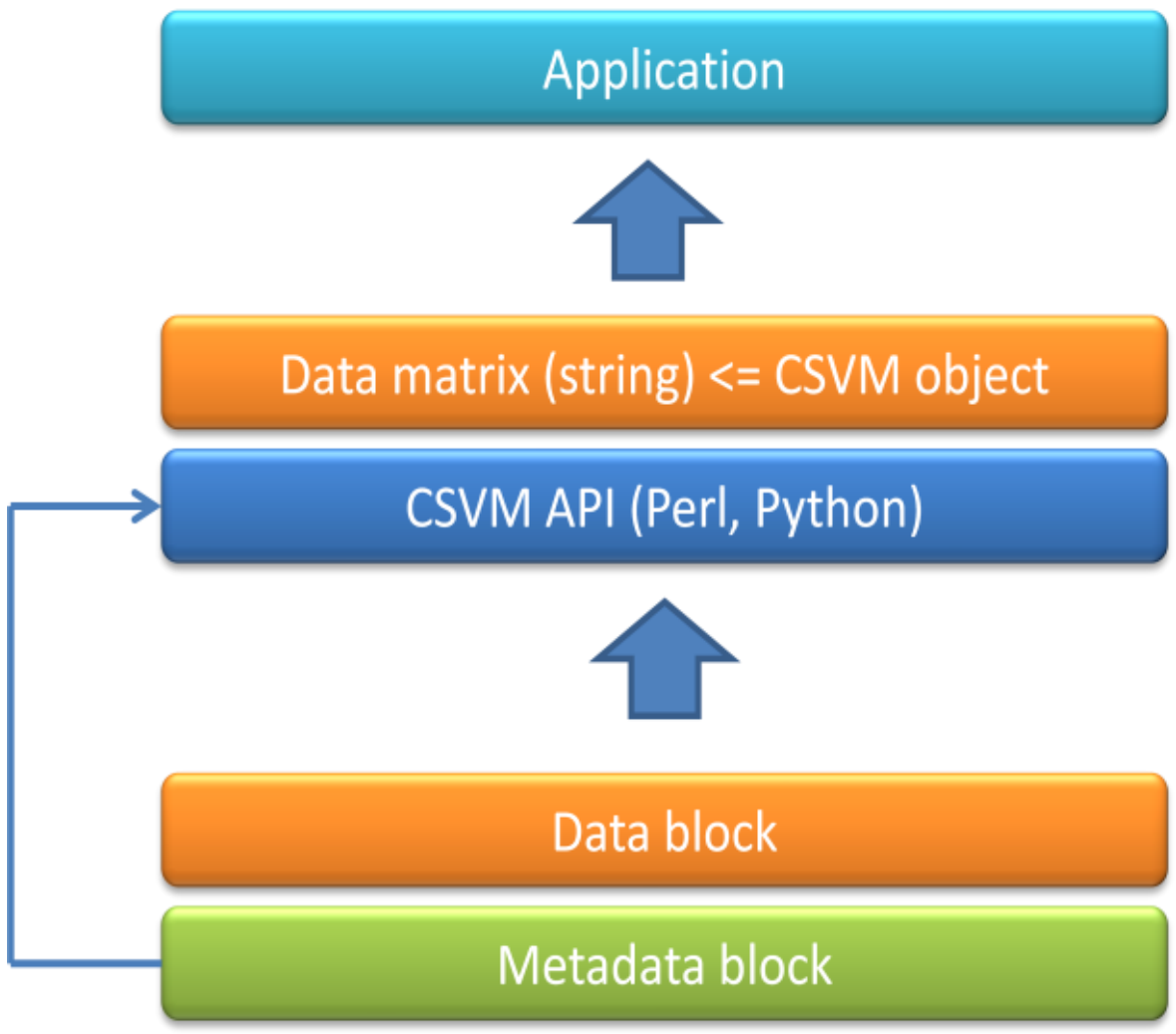

In Perl language, the CSVM data structure, used as a $csvm_ptr object is defined in build::csvm package and is constituted of the following fields:

*Table 12. – Perl CSVM records.*

| *Field* | *Type* | *Init* | *Information* |
|---|---|---|---|
| SOURCE | String | = undef; | CSV/CSVM file name |
| CSV | String | = undef; | CSV or CSVM (CSV filetype) |
| HEADER_N | Integer | = 0; | Value of c+1 |
| HEADER | Array [0..c] of String | = []; | CSVM #HEADER fields list |
| TITLE_N | Integer | = 0; | Typically 1, let for future use |
| TITLE | String | = undef; | CSVM title |
| TYPE_N | Integer | = 0; | Value of c+1 |
| TYPE | Array [0..c] of String | = []; | CSVM #TYPE fields list |
| WIDTH_N | Integer | = 0; | Value of c+1 |
| WIDTH | Array [0..c] of String | = []; | CSVM #WIDTH fields list |
| DATA_R | Integer | = 0; | r+1, numer of CSVM data rows |
| DATA_C | Integer | = 0; | c+1, numer of CSVM data columns |
| DATA | Matrix [0..r][0..c] of String | = []; | CSVM data matrix |
| META | String | = undef; | CSVM #META field string |

Here is a short explanation on how it is possible to parse a CSVM file, first of all you need to import the build::csvm Perl package :

```
use build::csvm;
```

After you need to define a csvm object, implemented as a pointer to a CSVM structure :

```
my $in_ptr = &csvm_ptr_new; ## pointer to csvm structure
```

At least you can parse a CSVM file in a CSVM structure, the string $in_file_path.$in_file_name is the path and name of CSVM file appended:

```
$in_ptr->csvm_ptr_read_csvm($in_file_path.$in_file_name);
```

If field delimiters are not TABs, you can use the method csvm_ptr_read_extended_csvm to load file with field ($sep value) specification:

```
$csvm_ptr->csvm_ptr_read_extended_csvm($fsep);
```

You can also load a file in a string ($s) transform the string and import the string in CSVM structure knowing line separator ($lsep) and field separator ($fsep) using the method csvm_ptr_get_csvm.

```
$csvm_ptr->csvm_ptr_get_csvm($s, $lsep , $fsep);
```

You can now apply some methods or access and transform each field of structure:

```
$in_ptr->csvm_ptr_dump(0,0); # echoes structure
$in_ptr->{'TITLE'} = "Modified title."; # change field title
print "[".$self->{'DATA'}->[$i][$j]."]"."\n";
```

In order to clean memory after all operations, we recommend the following sequence :

```
$in_ptr->csvm_ptr_clear;
undef $in_ptr;
```

It is possible to access to CSVM data without using a CSVM object, a subroutine is provided in build::csvm which returns a string matrix from CSVM file (the file contents are stored in a string $s):

```
my @matrix = &csvm_matrix_data($s, $lsep , $fsep , $fields_n);
```

In which $lsep and $fsep are the row and fields separators. Note that no analysis of CSVM string is done before parsing data in CSVM structure. In this case you need to provide the number of columns wanted in the resulting matrix (must see build::csvm &csvm_matrix_data for more information).

*The core of basic functionalities of the Perl toolkit is restricted in regard to Python toolkit, but shares some common functions and nearly the same object. The CSVM Perl toolkit (created 2002-2004) is now deprecated. Please contact corresponding author for support on this Perl toolkit.*

# 4. Using CSVM with Python programming language

The following lines are part of definition of csvm_ptr object defined in build.parsers.csvm package, with nearly the same fields found in Perl version :

*Figure 4. –Python CSVM object.*

```
class csvm_ptr:
   """
   Follows CSVM specs (v:1.x) for contents of data structure. Standard column
   types are NUMERIC,TEXT,DATE,BOOLEAN. Some of us, use also INTEGER, FLOAT
   for numeric types. Some of us, use also NODE, LINK, IMAGE for web data
   embedded in CSVM files. WIDTHs (10,50 if not set) are for Javascript tables
   and can be omitted.
   *** 1.01/080304/fred
   """
   def __init__(self):
       self.SOURCE = ""      # path/file name of readed CSVM file
       self.CSV = ""         # CSVM or CSV depending of file contents
       self.TITLE_N = 0      #Titles of CSVM file (let for future, only one string used today)
       self.TITLE = ""       # Title of CSVM file
       self.HEADER_N = 0     # Number of data columns titles
       self.HEADER = []      # List of data column titles
       self.TYPE_N = 0       # Number of data columns types (= self.HEADER_N)
       self.TYPE = []        # List of data column types
       self.WIDTH_N = 0      # Number of data columns widths (= self.HEADER_N)
       self.WIDTH = []       # List of data column widths
       self.DATA_R = 0       # Number of data rows
       self.DATA_C = 0       # Number of data columns (= self.HEADER_N)
       self.DATA = []        # String matrix containing data
       self.META = ""        # Meta string
```

## 4.1 Using the CSVM API

The description of CSVM APIs is out of scope of this document but it could be useful to provide an example to understand how a CSVM file could be used. Given the following pieces of code in Python, first we use a blank CSVM object :

```
   print "*** A new blank CSVM structure"
   c = csvm_ptr()
   print "*** Print the empty structure ... "
   c.csvm_ptr_dump(0,0)
   print "=> Is the empty structure is a CSVM object = ", csvm_iscsvm(c)
   print "=> Is self.DATA matrix is a CSVM object = ", csvm_iscsvm(c.DATA)
```

Which gives :

```
*** A new blank CSVM structure
*** Print the empty structure ...

DUMP: CSVM info {
SOURCE
CSV
META    []
TITLE_N 0
TITLE
HEADER_N       0
TYPE_N  0
WIDTH_N 0
DATA_R  0
DATA_C  0
        -1      -1

}
 done

=> Is the empty structure is a CSVM object =  True
=> Is self.DATA matrix is a CSVM object =  False
```

Now we will load a CSVM file, dumps it, dump a row :

```
print "*** Read test.csvm and fills the structure ..."
c = csvm_ptr_read_extended_csvm(c,"test/test1.csvm","\t")
print "*** Dump all ... "
c.csvm_ptr_dump(0,0)
print
print "*** Print as string the second row (numbered as 1 in CSVM) ... "
s = c.csvm_ptr_str_dump(2,0)
print s
```

The dump methods, printouts all metadata, for all the rows :

```
*** Read test.csvm and fills the structure ...
*** Dump all ...

DUMP: CSVM info {
SOURCE  test/test1.csvm
CSV     CSVM
META    [Test of|meta|fields|use]
TITLE_N 1
TITLE   CSV File [ test\test.csv ]
HEADER_N        15
TYPE_N  15
WIDTH_N 15
0       50      NUMERIC {numero}
1       50      TEXT    {fichier_mol}
2       50      TEXT    {nom}
3       50      NUMERIC {vrac}
4       50      TEXT    {plaque}
5       50      TEXT    {chimiste}
6       50      TEXT    {observations}
7       50      TEXT    {ref_produit}
8       50      TEXT    {ref_cahier}
9       50      TEXT    {code_labo}
10      50      TEXT    {no_equipe}
11      50      TEXT    {no_boite}
12      50      TEXT    {droits}
13      50      TEXT    {let_ligne_boite}
14      50      TEXT    {no_col_boite}
DATA_R  6
DATA_C  15
        6       15
0       [1][af01.mol][Tyrosine][10][oui][M.Dupont][existe sous forme de sel de
sodium][af01][C1][CCC][1][1][L][A][1]
1       [5][af02.mol][Histidine][20][oui][J.Smith][][af02][C1][CCC][1][1][L][B][1]
2       [2][af03.mol][Tryptophane][20][oui][nous][][af03][C1][CCC][1][1][L][C][1]
3       [3][af04.mol][Proline][12][non][eux][][af04][C2][][][][][][]
4       [4][af05.mol][Adenosine][0][oui][elle@ici][Plus de produit
disponible][af05][C1][CCC][1][1][L][F][3]
5       [6][af06.mol][Phosphatidyl Choline][300][non][lui@labas][Purifié a partir de jaune
d'oeuf][af06][D2][][][][][][]
}
 done
```

Or one particular row :

```
*** Print as string the second row (numbered as 1 in CSVM) ...

DUMP: CSVM info {SOURCE       test/test1.csvm
CSV     CSVM
META    [Test of|meta|fields|use]
TITLE_N 1
TITLE   CSV File [ test\test.csv ]
HEADER_N        15
TYPE_N  15
WIDTH_N 15
0       50      NUMERIC {numero}
1       50      TEXT    {fichier_mol}
2       50      TEXT    {nom}
3       50      NUMERIC {vrac}
4       50      TEXT    {plaque}
5       50      TEXT    {chimiste}
6       50      TEXT    {observations}
7       50      TEXT    {ref_produit}
8       50      TEXT    {ref_cahier}
```

```
9       50      TEXT    {code_labo}
10      50      TEXT    {no_equipe}
11      50      TEXT    {no_boite}
12      50      TEXT    {droits}
13      50      TEXT    {let_ligne_boite}
14      50      TEXT    {no_col_boite}
DATA_R  6
DATA_C  15
[5]     [af02.mol]      [Histidine]     [20]    [oui]   [J.Smith]       []      [af02]  [C1]    [CCC]   [1]
        [1]     [L]     [B]     [1]
 done
```

It easy to extract one column using its index value, as a string with a new delimiter or a classical Python list of strings:

```
print "*** Extract the second column (numbered as 1 in CSVM)"
print "-> as string with a separator '|'"
print csvm_ptr_str_getvec(c,0,2,"|")
print "-> as vector (Python list) of strings"
print csvm_ptr_getvec(c, 0, 2)
```

```
*** Extract the second column (numbered as 1 in CSVM)
-> as string with a separator '|'
af01.mol|af02.mol|af03.mol|af04.mol|af05.mol|af06.mol
-> as vector (Python list) of strings
['af01.mol', 'af02.mol', 'af03.mol', 'af04.mol', 'af05.mol', 'af06.mol']
```

But it is also possible to extract a column using the value of #HEADER (column titles, strict_mode=1) or a substring in this value (strict_mode = 0):

```
print "*** Extract columns on the value of headers"
print "-> the column named 'fichier_mol' in strict mode"
ls = csvm_ptr_getcol(c, 'fichier_mol', 1)
print "found %d column in CSVM stream" % (ls[0])
print ls[1]
print "-> the columns in which string 'no_' is found"
ls = csvm_ptr_getcol(c, 'no_', 0)
print "found %d columns in CSVM stream" % (ls[0])
for i in range (1, ls[0]+1, 1):
    print i, ls[i]
```

```
*** Extract columns on the value of headers
-> the column named 'fichier_mol' in strict mode
found 1 column in CSVM stream
['af01.mol', 'af02.mol', 'af03.mol', 'af04.mol', 'af05.mol', 'af06.mol']
-> the columns in which string 'no_' is found
found 3 columns in CSVM stream
1 ['1', '1', '1', '', '1', '']
2 ['1', '1', '1', '', '1', '']
3 ['1', '1', '1', '', '3', '']
```

Then changing some this such as the delimiter is very simple and we could make a new CSVM file, using the § as new delimiter.

```
print "*** Making a new CSVM as string for export (changing th field separator)"
s = csvm_ptr_make_csvm(c,"\n","§")
print s
```

```
*** Making a new CSVM as string for export (changing the field separator)
1§af01.mol§Tyrosine§10§oui§M.Dupont§existe sous forme de sel de sodium§af01§C1§CCC§1§1§L§A§1§
5§af02.mol§Histidine§20§oui§J.Smith§§af02§C1§CCC§1§1§L§B§1§
2§af03.mol§Tryptophane§20§oui§nous§§af03§C1§CCC§1§1§L§C§1§
3§af04.mol§Proline§12§non§eux§§af04§C2§§§§§§§
4§af05.mol§Adenosine§0§oui§elle@ici§Plus de produit  disponible§af05§C1§CCC§1§1§L§F§3§
6§af06.mol§Phosphatidyl Choline§300§non§lui@labas§Purifié a partir de jaune d'oeuf§af06§D2§§§§§§§

#TITLE§CSV File [ test\test.csv ]
#HEADER§numero§fichier_mol§nom§vrac§plaque§chimiste§observations§ref_produit§ref_cahier§code_labo§no
_equipe§no_boite§droits§let_ligne_boite§no_col_boite
#TYPE§NUMERIC§TEXT§TEXT§NUMERIC§TEXT§TEXT§TEXT§TEXT§TEXT§TEXT§TEXT§TEXT§TEXT§TEXT§TEXT
#WIDTH§50§50§50§50§50§50§50§50§50§50§50§50§50§50§50
#META§Test of|meta|fields|use
```

It is also simple to export as CSV flow, notice that a Python module exists in API to import CSV files as they were CSVM files.

```
print "*** Making a simple CSV as string for export (changing the field separator)"
s = csvm_ptr_make_csv(c,"\n","§",'"')
print s
```

```
*** Making a simple CSV as string for export (changing th field separator)
"numero"§"fichier_mol"§"nom"§"vrac"§"plaque"§"chimiste"§"observations"§"ref_produit"§"ref_cahier"§"c
ode_labo"§"no_equipe"§"no_boite"§"droits"§"let_ligne_boite"§"no_col_boite"
1§"af01.mol"§"Tyrosine"§10§"oui"§"M.Dupont"§"existe sous forme de sel de
sodium"§"af01"§"C1"§"CCC"§"1"§"1"§"L"§"A"§"1"
5§"af02.mol"§"Histidine"§20§"oui"§"J.Smith"§""§"af02"§"C1"§"CCC"§"1"§"1"§"L"§"B"§"1"
2§"af03.mol"§"Tryptophane"§20§"oui"§"nous"§""§"af03"§"C1"§"CCC"§"1"§"1"§"L"§"C"§"1"
3§"af04.mol"§"Proline"§12§"non"§"eux"§""§"af04"§"C2"§""§""§""§""§""§""
4§"af05.mol"§"Adenosine"§0§"oui"§"elle@ici"§"Plus de produit
disponible"§"af05"§"C1"§"CCC"§"1"§"1"§"L"§"F"§"3"
6§"af06.mol"§"Phosphatidyl Choline"§300§"non"§"lui@labas"§"Purifié a partir de jaune
d'oeuf"§"af06"§"D2"§""§""§""§""§""§""
```

The following call could be used to close csvm_ptr object and free the memory.

```
print "*** Close CSVM structure."
c.csvm_ptr_clear()
```

*This is a limited set of Python toolkit that are shown here.*
*Please contact corresponding author for support on the Python (Pybuild) toolkit.*

## 5. The CSVM parser

We have defined fuzzy data types such as TEXT or NUMERIC as a signal to users and code designers, because from our point of view, CSVM format (particularly #TYPE row) must not be constrained by a data type interpretation and a normative list of allowed types.
Reading the previous examples, it is obvious that the CSVM parser doesn't interpret the data using #TYPE keyword: all is parsed as strings (scalar, lists and matrix) in memory. The parser is uncoupled from data, this gives the ability to use any kind of data types in the CSVM file. The application layer must interpret the information depending on the metadata stored in the CSVM file.
Now, what about if *n* programmers use *m* different types for the same data/data type, in *k* CSVM files? Even if CSVM is a trivial concept, it provides simple mechanisms to go over these kinds of problems. We take an example: some CSVM files (*k*A) with particular types that are documented by one CSVM file (B). The B-File could have the following contents:

*Table 13. – A simple CSVM dictionary.*

```
CONCENTRATION   Mol/L   NUMERIC   FLOAT
MOL_CONC        MOL/L   NUMERIC   FLOAT
MASSIC_CONC     G/L     NUMERIC   FLOAT
MOLECULE_ID     -       NUMERIC   INTEGER
NAME            -       TEXT      STRING
FORMULA         -       TEXT      SMILES
FORMIMG         -       IMAGE     URL_SMALL|URL_IMAGE

#TITLE    Small types table
#HEADER   NAME  UNIT  CSVM_TYPE   REAL TYPE
#TYPE     TEXT  TEXT  TEXT        TEXT
#WIDTH    10    10    10          10
#META     Version 1.1 (Jule 12 2011) by [John]
```

In this file, we store the #HEADER values of file A, eventually the units, and a real type to be used by application. If some coworkers had used the same information in CSVM files, but with different names (i.e. CONCENTRATION vs. MOL_CONC) or different units (MASSIC_CONC vs. MOL_CONC) the data of B-File could be used to restore initial information or normalize data.
Some less obvious types can be defined (i.e. a molecular formula expressed with SMILES[5] format). Composite format can also be define, in the last row an image is defined by an URL for the reduced picture and another for the full picture.
With the same parser and the same file format we can read the two CSVM files, these with data and the one with information on the previous. This kind of approach can be used also to define conversion of metadata/data (typically units, names, types …) between CSVM files.
If this kind of auxiliary file is used for conversion/normalization, we have defined a first extension of CSVM files (readable with the same parser) and called it CSVM dictionaries. This concept is very similar to the B-file but can handle also type conversions for CSVM metadata without changing file format.

## 6. Open Format

Due to metadata block added to CSV block, CSVM is not RFC4180[6] compliant.
CSVM is as an Open Format following the definition proposed by the Linux Information Project [7] (any format that is published for anyone to read and study but which MAY OR MAY NOT BE encumbered by patents, copyrights or other restrictions on use). We expect that CSVM could be a Free Format (IS NOT encumbered by any copyrights, patents, trademarks or other restrictions) after a running time.

---

5 Chemical File Format [Wikipedia] - http://en.wikipedia.org/wiki/Chemical_file_format
6 Y. Shafranovich (2005), Common Format and MIME Type for Comma-Separated Values (CSV) Files, RFC 4180, The Internet Society - http://tools.ietf.org/html/rfc4180
7 Open Format [Wikipedia] - http://en.wikipedia.org/wiki/Open_format

## 7. Conclusion and perspectives

Different fields of science were sampled, and computer scientists have used CSVM files for a range of current tasks needed by laboratory's activities:

- Pure tabular data, punching card type data;
- Indexes, collection of files;
- External parameters files for software components (wrappers, converters, filters …);
- Aggregation of tables;
- Conversion of table data (CSVM dictionaries);
- Mixed file for data, parameters, results (times series);
- Driver for software pipes (linear, parallel);
- Key / value(s) files;
- Database schema and tables
- Trees.

We expect that this list is not exhaustive, because the typical range of CSVM files is from 1-1000, 5000 rows and 1-20, 30 columns, about the same range than a current spreadsheet file. CSVM is useful to store data in long term and to be included in software pipes, particularly with automation written in Python.
CSVM is not a substitute for formats in use to handle mass data (HF5, NetCDF …) with ASCII or binary storage. CSVM is not also a substitute of XML, because it is focused on tabular data, and a same generic parser could be used for all kind of data and data types. CSVM is a light format designed to handle complexity of small data islands handled every day.
Our main effort is focused on extension of CSVM leading to CSVM-2 specification. We want take in account a lot of requirements not easy fulfilled by other light data formats, particularly: 1) to embed ASCII or binary data in the same ASCII table, 2) to embed CSVM data inside a CSVM cell and make rooted trees of mixed CSVM tables and single CSVM cells.

## References

- G. Beyries. Composants logiciels génériques pour les collections de données. Mémoire ingénieur Ecole des Techniques du Génie Logiciel (2004).
- G. Beyries, F. Rodriguez. Quels outils informatiques pour les collections de données ? Restitution Programme ECOBAG P1, Agen (2004).
- Software design and components for enzymology, S.Gavalda, F.Rodriguez, G.Beyries, C.Blonski. 21ème CBSO, Ax-Les-Thermes (2006).
- M. Gerino et coll. Bilan et dynamique de la matière organique et des contaminants au sein d'une discontinuité ex de la retenue de Malause. Restitution des travaux scientifiques du projet de recherche Ecobag P2 (2006).
- S. Gau, F. Rodriguez, C. Lherbet, C. Menendez, M. Baltas. Approches interdisciplinaires pour la conception rationnelle d'inhibiteurs. RECOB 13, Aussois (2010).

## Acknowledgments

We are grateful to Professor Pierre Tisnes (LSPCMIB, UMR 5068 CNRS-Toulouse University) and Dr. Philippe Vervier (CNRS-Toulouse University and GIS ECOBAG) for supporting this first phase of work.
We thank Pr. Magali Gerino (ECOLAB, UMR 5245 CNRS-Toulouse University) and collaborators for advices and helpful discussions about this work.
We would like to thank the CNRS, the "Université Paul Sabatier" and GIS ECOBAG (2002-2004) for their financial support.
We thank all collaborators in different laboratories which shared data with us and helped us to develop the format's usage